\documentclass[11pt,twoside]{article}
\usepackage{amsmath,amssymb,fancyhdr,paralist}
\usepackage[perpage]{footmisc}

\makeatletter
\let\ps@plain\ps@fancy
\makeatother

\makeatletter
\def\@xfootnote[#1]{%
  \protected@xdef\@thefnmark{#1}%
  \@footnotemark\@footnotetext}
\makeatother

\newcommand{\stars}[0]{~\par{\centering{\bf *$\qquad$*$\qquad$*}~\par}~\par}

\begin{document}

\setcounter{page}{966}

\title{\vspace{-2em}On the Unification Problem in Physics\footnote{\scriptsize {\em Zum Unit\"atsproblem der Physik}, Sitzungsber. Preuss. Akad. Wiss. Berlin (Math. Phys.) 1921, 966--972}}
\author{from \textsc{Th.~Kaluza\footnote{\scriptsize Revised translation by V. T. Toth, based in part on translation by T. Muta, HUPD-8401, March 1984, Dep. Phys. Univ. Hiroshima.}}\\{\small in K\"onigsberg}}
\date{\small (Submitted by Mr. Einstein on December 8, 1921; s. above, p. 859.)}

\maketitle

\vskip -1.25em

In the general theory of relativity, in order to characterize
world events, the fundamental metric tensor $g_{\mu\nu}$ of the 4-dimensional world manifold,
interpreted as the the tensor potential of gravitation must be introduced
separately from the electromagnetic four-potential $q_\mu$.

The dualistic nature of gravitation and electricity still
remaining here does not actually destroy the ensnaring beauty
of either theory but rather affords a new challenge towards their
triumph through an entirely unified picture of the world.

A few years ago \textsc{H. Weyl}\footnote{Sitzungber. d. Berl. Acad. 1918 p.465.} put forward a surprisingly courageous
attack to the solution of these problems which belong among the
most magnificent ideas of the human spirit. Through the radical reexamination
of the geometric grounds he obtained, in addition to the
tensor $g_{\mu\nu}$, a kind of fundamental metric vector and interpreted
it as the electromagnetic potential $q_\mu$: There the complete
world metric is claimed to be the common source of all the natural
phenomena.

Here we strive for the same goal in a different way.

If we disregard the difficulty which is associated with the
practical application of \textsc{H. Weyl}'s profound theory, it is thought to
be quite idealistically possible to realize completely the unification
idea in which the gravitational and electromagnetic
fields stem from a single universal tensor. --- I would now like to
show that such an intimate combination of the two forces in the
world appears to be possible in principle.

\vskip -0.75em
\stars\vskip -0.75em

The rotational form of the electromagnetic field components
$F_{\varkappa\lambda}$ and also the apparent formal equivalence in the structure of
the gravitational and electromagnetic equations\footnote{In this respect refer to \textsc{H. Thirring}. Phys. Ztschr. 19 p.204.} lead us to the
conjecture that $\dfrac{1}{2}F_{\varkappa\lambda}=\dfrac{1}{2}(q_{\varkappa,\lambda}-q_{\lambda,\varkappa})$\footnote{By indices divided by a comma is meant the differentiation with respect to the corresponding world-parameter.}
may somehow be equal to the truncated three-index quantities $\left[\displaystyle\begin{matrix}i\lambda\\\varkappa\end{matrix}\right]=\dfrac{1}{2}(g_{i\varkappa,\lambda}+g_{\varkappa\lambda,i}-g_{i\lambda,\varkappa})$.
Provided there is a room for this idea, one may be drawn with greater confidence
to a path that previously seemed less attractive: Since in a four-dimensional
world, beyond the three-index quantities used as the field components of gravitation,
no further such quantities exist,
such an interpretation of $F_{\varkappa\lambda}$ is hardly supported unless
one makes the otherwise extremely odd decision to ask for help from a
new fifth dimension of the world.

Although our rich physical experience obtained so far provides
little suggestion to such an extra world-parameter, we are certainly
free to consider our space-time to be a four-dimensional part
of $R_5$; one then has to take into account the fact that we are
only aware of the space-time variability of state parameters, by
making their derivatives with respect to the new parameter
vanish or by considering them to be small as they are of higher
order (``cylinder condition''). Misgivings about the retrogressive
introduction of the fifth dimension are rendered groundless by
connecting the world parameter to the three-index quantities.

\stars

We therefore enter into $R_5$ and extend \textsc{Einstein}'s ansatz to
$R_5$: at the same time, we introduce the new parameter $x^0$ in addition
to the usual $x^1$ to $x^4$. If the fundamental metric tensor
of this $R_5$ is expressed as $g_{rs}$\footnote{Latin indices always run from 0 to 4 and Greek ones only
from 1 to 4.}, the three-index quantities $\left[\displaystyle\begin{matrix}ik\\l\end{matrix}\right]$,
denoted here by $-\Gamma_{ikl}$, become, by virtue of the cylinder condition,
\begin{align}
2\Gamma_{\varkappa\lambda\mu}&=g_{\varkappa\lambda,\mu}-g_{\lambda\mu,\varkappa}-g_{\mu\varkappa,\lambda}\qquad\text{(as before)},\nonumber\\
2\Gamma_{0\varkappa\lambda}&=g_{0\varkappa,\lambda}-g_{0\lambda,\varkappa},\qquad 2\Gamma_{\varkappa\lambda 0}=-(g_{0\varkappa,\lambda}+g_{0\lambda,\varkappa}),\\
2\Gamma_{00\varkappa}&=g_{00,\varkappa},\qquad 2\Gamma_{0\varkappa 0}=-g_{00,\varkappa},\qquad 2\Gamma_{000}=0.\nonumber
\end{align}

This result is at first sight hardly encouraging: Indeed
$\Gamma_{0\varkappa\lambda}$ appear in the form of a rotation, but the ten $\Gamma_{\varkappa\lambda 0}$, which would
have to be of electric nature according to our interpretation,
are in danger of being in the way. Nevertheless, we further
investigate the above result and, in order to keep $\Gamma_{0\varkappa\lambda}$ proportional
to $F_{\varkappa\lambda}$ we set
\begin{align}
g_{0\varkappa}=2\alpha q_{\varkappa},&\quad g_{00}=2\mathfrak{g},
\end{align}
so that the fundamental metric tensor of $R_5$ substantially becomes
the gravitational tensor potential framed by the electromagnetic
four-potential: The role of the component $\mathfrak{g}$ in the corner remains
undetermined for the time being. Introducing the shorthand $\Sigma_{\varkappa\lambda}$ for the sum $q_{\varkappa,\lambda}+q_{\lambda,\varkappa}$ corresponding to $F_{\varkappa\lambda}$, we have
\vspace{-0.5em}
\begin{align}
\Gamma_{0\varkappa\lambda}=\alpha F_{\varkappa\lambda},\quad\Gamma_{\varkappa\lambda 0}=-\alpha\Sigma_{\varkappa\lambda},\quad\Gamma_{00\varkappa}=-\Gamma_{0\varkappa 0}=\mathfrak{g}_{,\varkappa}.
\label{eq:3}
\end{align}
Consequently the electromagnetic field $F_{\varkappa\lambda}$,
its ``associated'' field $\Sigma_{\varkappa\lambda}$ and the gradient\footnote{In the four-dimensional sense.} of $\mathfrak{g}$
use up the thirty-five new three-index symbols (five of which vanish). Moreover,
from the comprehensive equality
\vspace{-0.5em}
\begin{align}
(\Gamma_{ikl}+\Gamma_{kli}+\Gamma_{lik})_{,m}=\Gamma_{mik,l}+\Gamma_{mkl,i}+\Gamma_{mli,k}
\end{align}
arise the well-known relations by virtue of the cylinder condition:
\vspace{-0.5em}
\begin{align}
\tag{4a}
F_{\varkappa\lambda,\mu}+F_{\lambda\mu,\varkappa}+F_{\mu\varkappa,\lambda}=0\quad\text{and}\quad\mathfrak{g}_{,\varkappa\lambda}=\mathfrak{g}_{,\lambda\varkappa}.
\end{align}

\vspace{-1.5em}
\stars
\vspace{-1em}

We now restrict, as usual, the choice of parameters by
$g=|g_{rs}| = -1$ and let $g_{rs}$ differ only a little bit from the
``Euclidean'' value $-\delta_{rs}$ (approximation I). With $\Gamma_{ik}^l=-\left\{\displaystyle\begin{matrix}ik\\l\end{matrix}\right\}=-\Gamma_{ikl}$
the intereseting components of the two four-index tensors:
\begin{align}
\begin{matrix}
\{\varkappa\lambda,\mu 0\}=\alpha F_{\varkappa,\mu}^\lambda,\quad\{\varkappa 0,0\lambda\}=-\mathfrak{g}_{,\varkappa\lambda},\\
\{\varkappa\lambda,00\}=\{\varkappa 0,00\}=\{00,00\}=0.
\end{matrix}
\end{align}
Fortunately the associated field in Eq.~(\ref{eq:3}) does not participate
here: Among electric quantities only derivatives of the
fields appear which determine the curvature of $R_5$. Further, by
making the contracted tensor $R_{ik}=\{ir,rk\}$, we find, according
to our assumptions (in the well-known notation):
\begin{align}
R_{\mu\nu}&=\Gamma_{\mu\nu,\rho}^\rho,\qquad\text{(as earlier)},\nonumber\\
R_{0\mu}&=-\alpha\Delta\iota v_\mu F,\label{eq:6}\\
R_{00}&=-\Box\mathfrak{g}.\nonumber
\end{align}
Therefore the fifteen components of the curvature tensor
on the left-hand side reduce into:
\begin{inparaenum}
\item the ordinary field equations of gravitation,
\item the basic electromagnetic equations and
\item a \textsc{Poisson} equation for the yet unexplained $\mathfrak{g}$.
\end{inparaenum}
Therein lies the foremost justification
of our ansatz and the hope to consider gravitation and
electricity as manifestations of a universal field.

\vspace{-1em}
\stars
\vspace{-1em}

For the energy-momentum tensor of matter, dominating
the right-hand side of the field equations in $R_5$, holds the following
under approximation I:
\begin{equation}
T_{ik}=T^{ik}=\mu_0u^iu^k,\label{eq:7}
\end{equation}
$$(\mu_0=\text{rest mass density},\quad u^r=\frac{dx^r}{ds},\quad ds^2=g_{lm}dx^ldx^m);$$

Since now (for all three types of field equations) $R_{0\mu} = -\varkappa T_{0\mu}$,
it follows from the \textsc{Maxwell} equations that according to Eq.~(\ref{eq:6})
for the components of the four-current:
\begin{equation}
{\mathbf I}^\mu=\rho_0v^\mu=\frac{\varkappa}{\alpha}T_{0\mu}=\frac{\varkappa}{\alpha}\mu_0u^0u^\mu\label{eq:8}
\end{equation}
$$(\rho_0=\text{rest charge density},\quad v^\rho=\frac{dx^\rho}{d\sigma},\quad d\sigma^2=g_{\lambda\mu}dx^\lambda dx^\mu);$$
the space-time energy-momentum tensor is thus framed essentially by the current density.

We then continue further investigation under the assumption
$u^0$, $u^1$, $u^2$, $u^3\ll 1, u^4\sim 1$ (approximation II). This
implies not only a small velocity but also a very tiny specific
charge $\dfrac{\rho_0}{\mu_0}$ of moving matter; in fact, as thereupon $d\sigma^2\sim ds^2$, $v^\rho\sim u^\rho$,
it follows from Eq.~(\ref{eq:8}) if one sets\footnote{According to the equation of motion; see the next section.} $\alpha=\sqrt{\dfrac{\varkappa}{2}}=3.06\times 10^{-14}$:
\vspace{-1em}
\begin{align}
\rho_0=\frac{\varkappa}{\alpha}\mu_0u^0=2\alpha\mu_0u^0\ll\mu_0.\tag{8a}
\end{align}
First of all this equation teaches us again that in this case
we need to understand the electric charge essentially as the
fifth component of the energy-momentum of the matter ``moving
across'' the space $x^0=\rm{const}$.: A further merger of two formerly
heterogeneous basic concepts thus appears.

Because finally $T_{00}, T_{11}, T_{22}, T_{33}\sim 0$ in approximation II,
we find according to Eq.~(\ref{eq:7}):
\begin{align}
T=g^{ik}T_{ik}=-T_{44}=-\mu_0,
\end{align}
so that for the ordinary form of the field equation of the first kind:
\begin{align}
R_{00}=-R_{44}=\frac{\varkappa}{2}\mu_0.
\end{align}
The corner potential $\mathfrak{g}$ is also proven, due to Eq.~(\ref{eq:6}), to be
essentially minus the gravitational potential, while $\mathfrak{G} = \dfrac{g_{44}}{2}$
keeps the former meaning.

\stars

After we thus disposed of the standard quantities in the
field equations in a satisfactory way, we still face with the
question whether the ``geodesic'' equation of motion in $R_5$,
\begin{align}
\dot{u}^l=\frac{du^l}{ds}=\Gamma^l_{rs}u^ru^s\label{eq:11}
\end{align}
then represents the motion of charged matter in the gravitational
and electromagnetic field in accord with experiments. In approximation
II this is immediately the case: Due to the interchangeability
of $ds$ and $d\sigma$ one obtains according to Eq.~(\ref{eq:3}):
\begin{align}
\bar{v}^\lambda=\frac{dv^\lambda}{d\sigma}=\Gamma^\lambda_{\rho\sigma}v^\rho v^\sigma+2\alpha F^\lambda_\varkappa u^0u^\varkappa-\mathfrak{g}_{,\lambda}{u^0}^2,\tag{11a}
\end{align}
i.e., due to the smallness of the term with ${u^0}^2$ for the force
density one finds
\begin{align}
\pi^\lambda=\mu_0\bar{v}^\lambda=\Gamma_{\rho\sigma}^\lambda T^{\rho\sigma}+F_\varkappa^\lambda\mathbf{I}^\varkappa\quad\left(\alpha=\sqrt{\frac{\varkappa}{2}}\quad\text{adapted; see footn.}\right).
\end{align}
The total force thus splits automatically into a gravitational
and electromagnetic part of the ordinary form.

Finally, for the 0-component of Eq. (\ref{eq:11}) there remains
only
\begin{align}
\dot{u}^0=\alpha\Sigma_{44}=2\alpha q_{4,4},\tag{11b}
\end{align}
so that, in approximation II, the quasi-static
$\dfrac{d}{dx^4}\left(\dfrac{\rho_0}{\mu_0}\right)=2\varkappa q_{4,4}$\footnote{Cf. (8a).} becomes small in higher orders: The
necessary constancy of $\rho_0$ seems to be guaranteed accordingly.

Hence also for the equations of motion the associated field
remains insignificant in our approximation.

\stars
\newpage

If approximation II was in accord with the reality,
the unification theory under investigation would be executed,
as mentioned earlier, substantially in a satisfactory manner:
A unique potential tensor generates a universal field which splits,
in ordinary condition, into a gravitational and electromagnetic
part.

Now, however, the matter at least in its smallest building
blocks is not at all weakly charged: following \textsc{Weyl}'s expression,
its ``macroscopic rest'' is confronted with its ``microscopic
restlessness'', and this holds, in the above-mentioned view, quite
specifically for the new world-parameter $x^0$: For the electron
or H-nucleus, $\dfrac{\rho_0}{\mu_0}$ is not small and so the ``velocity'' component
$u^0$ is not at all small! In the form constrained by approximation
II the theory can at best coarsely describe macroscopic phenomena,
and a fundamental problem arises concerning its very applicability to
those elementary particles.

If one now, nevertheless, tries to describe the motion of the
electron by geodesics in $R_5$, one immediately meets with the serious
difficulty\footnote{I wish to thank Mr. Einstein for his valuable interest in
the origin of the above assumptions and the suggestion of
the inconsistencies described here.} which threatens to destroy the foundation laid here.
It lies simply in the fact that, with the rigid application
of the previous assumptions, $u^0$ for the electrons becomes
enormously large owing to $\dfrac{e}{m} = 1.77\times 10^7$ (reduced in light
seconds) so that the last term in Eq.~(lla), instead of
vanishing, assumes a value that exceeds everything and defies all
experience, provided that everything remains formally as before.
Now the larger value of $u^0$ indeed implies modifications of the
theory anyhow (the interchangeability of $ds$ and $d\sigma$ is lost), but
it seems almost impossible to develop the theory only in the old
framework without any new hypotheses.

Nevertheless -- with full reservations -- I believe in finding,
clearly in the following direction, a way which provides a fully
satisfactory point of view leading to the goal. Since
$R_{00}\sim -R_{44}$ even for arbitrary $u^0$ provided that the velocity of
the matter which generates fields is not too large, two gravitational
terms in Eq.~(lla) take opposite signs if the nature of $x^0$
which has been completely irrelevant is determined suitably.
And it then seems that, apart from the gravitational constant $\varkappa$
which is slightly questionable anyhow, a reconciliation is
possible between two compelling magnitudes for which gravitation
is obtained as a kind of the difference effect. One is tempted
to be bribed by this possibility with the prospect that the role
of a statistical quantity may be assigned to that constant. At
the moment the consequence of this hypotheses is not yet examined
sufficiently; other possibilities should also be searched for.
After all, what threatens all the ansatz which demand universal
validity is the sphinx of modern physics -- quantum theory.

\stars

In spite of full recognition of the aforementioned physical
as well as epistemological difficulties piled up in front of our
understanding developing here, it is rather difficult for us to
believe that, in all those relations which in their formal unity
seem hard to be replaced by anything else, only a capricious
accident performs its alluring play. If it could be proven some
day that there exists more behind the presumed relations than
merely meaningless formalism, then this would certainly imply
a new triumph for \textsc{Einstein}'s general theory of relativity, whose
appropriate application to the five-dimensional world comes into
question.

~\par

%Issued on 30th of January in 1922.

%Berlin, printed by the Empire Printing Co.

\end{document}